\documentstyle[prl,aps,amssymb]{revtex}

\tighten

\begin{document}
\draft
\title{Fragility of Thermodynamically-Abnormal Ground States of 
Finite Systems}
\author{Takayuki Miyadera\cite{mi} and Akira Shimizu\cite{shmz}}
\address{
Department of Basic Science, University of Tokyo, 
3-8-1 Komaba, Meguro-ku, Tokyo 153-8902, Japan
}
\date{\today}
\maketitle

\begin{abstract}
We study a general macroscopic quantum system of a finite size, 
which will exhibit a symmetry breaking if the system size goes 
to infinity, when the system interacts with an environment.
We evaluate the decoherence rates of the anomalously fluctuating 
vacuum (AFV), which is the symmetric ground state, and the pure 
phase vacua (PPVs).  By making full use of the locality and huge 
degrees of freedom, we show that there can exist an interaction 
with an environment which makes the decoherence rate of the AFV 
anomalously fast, whereas PPVs are less fragile.
\end{abstract}
We consider a macroscopic quantum system which can exhibit 
spontaneous symmetry breaking (SSB).
According to experience, 
it is very hard to observe superpositions of 
states with different values of the order parameter.
We call the ground state of such pure states 
the {\it anomalously fluctuating vacuum} (AFV) because
it has anomalously large fluctuations 
of macroscopic variables.
When the system volume is infinite, 
the reason for the impossibility of observing 
the AFV is obvious; there is no local operator intertwining 
the macroscopically distinct states, 
and thus their superposition is a mixed state
rather than a pure state 
\cite{ruelle,haag}.
However, this reasoning cannot be applied to 
finite systems, and any superposition of pure states 
is a pure state (except when a superselection rule forbids it)
in quantum theory of finite {\em closed} systems.
Hence, the answer should come from the fact that real physical systems are 
not completely closed; there are interactions with 
surrounding environments.
Effects of environments have been discussed intensively 
in studies of, e.g.,  `macroscopic quantum coherence' \cite{AL}
and quantum measurement \cite{Zurek}.
However, most previous studies on these subjects assumed
that the principal systems of interest were describable 
by a {\it small number} of collective coordinates,
which interact {\it non-locally} with some specific environment.
Although such models might be applicable to systems which 
have a non-negligible 
energy gap to excite `internal coordinates' of 
the collective coordinates, 
there are many systems which do not have such an energy gap.
Moreover, 
the results depended strongly on the choices of 
the coordinates and the form of the nonlocal interactions, 
so that general conclusions were hard to draw.

In this work, we study a
{\it general} finite system of {\em huge} degrees of freedom $|\Lambda|$, 
which interacts with a {\em general} environment E
via a general {\em local} interaction $H_{\rm int}$.
We derive a lower bound $\gamma$ of the decoherence rates
for the AFV and for `pure phase vacua' (PPVs), which 
will be defined later, 
by making full use of the locality:
the interaction must be local (Eq.\ (\ref{eqn:int}) below) 
and macroscopic variables must be 
averages over a macroscopic region (Eq.\ (\ref{intensiveOp})).
To express the locality manifestly, 
we use a local field theory throughout this work.
It is shown that there can exist $H_{\rm int}$ which makes 
$\gamma$ of the AFV larger than that of PPVs by an
anomalously large factor ${\cal O}(|\Lambda|)$. 
We also derive a lower bound $\Delta \gamma$ of 
the difference of the decoherence 
rates between the AFV and PPVs, 
and show that there can exist
$H_{\rm int}$ which makes 
$\Delta \gamma$ anomalously large, 
proportional to ${\cal O}(|\Lambda_{\rm C}|)$, 
where $\Lambda_{\rm C}$ ($\subseteq \Lambda$) is the 
`contact region' in which 
the principal system interacts with E.
These results show that 
the AFV is `fragile' (i.e., decoheres at an anomalously fast 
rate) for large $|\Lambda|$ and $|\Lambda_{\rm C}|$,
however small the coupling constant of $H_{\rm int}$ is,
whereas PPVs are less fragile.


We first fix the energy scale $\Delta E$ of interest. 
Since it sets 
a minimum length scale $l$, 
we can treat the system as a lattice system 
$\Lambda$ whose lattice constant is $l$.
In some cases, 
the degrees of freedom (the number of lattice cites)
$|\Lambda|$ of the effective theory can become small
even for a macroscopic system
when, e.g., a non-negligible energy gap exists in $\Delta E$,
so that the number of quantum states in $\Delta E$ is small.
Some SQUID systems are such examples.
We here exclude such systems, and 
concentrate on systems whose $|\Lambda|$ is a macroscopic number.
%
Although $l$ is somewhat arbitrary, this ambiguity 
does not change the conclusions of the present paper.
We take $l = 1$, and 
consider the case where $\Lambda$ is 
a $d$-dimensional hypercubic lattice system $L^d$.
%
For simplicity, we impose the periodic boundary conditions, 
and assume that
all states under consideration are invariant under the spatial translation.
To establish the relation between infinite systems
and macroscopic but finite systems of our interest, 
we consider a sequence of lattice systems $\{ \Lambda  \}$ 
with increasing $|\Lambda|$.
We assume that it exhibits an SSB as $\Lambda \to {\bf Z}^d$, 
for which a ground state $\Xi_{{\bf Z}^d}$  
is a pure phase vacuum (PPV) that breaks the symmetry, 
i.e.,
the expectation value 
of some order parameter $m(x)$, which is a one-site observable 
at site $x$,
is nonvanishing.
For a finite $| \Lambda |$, 
we can find a sequence of pure states 
$\{ \Xi_{\Lambda}\}$ that approaches 
$\Xi_{{\bf Z}^d}$
as $\Lambda \to {\bf Z}^d$ \cite{HL,miyashita,KT,SMpre}.
If we take $\Xi_{\Lambda}$ as a normalized vector
in a Hilbert space on $\Lambda$, it satisfies
$
\nu_{\Lambda} 
\equiv(\Xi_{\Lambda},m(0) \Xi_{\Lambda}) \to \nu_{{\bf Z}^d} \neq 0
$
as $\Lambda \to {\bf Z}^d$.
Note that $\Xi_{{\bf Z}^d}$ has the ``cluster property'' 
\cite{ruelle,haag},
which means that spatial correlations of any local operators 
vanish at a large distance.
To exclude exceptional uninteresting sequences, 
we require that 
$\Xi_{\Lambda}$'s should also have the cluster property,
and call such $\{\Xi_{\Lambda}\}$, as well as its element $\Xi_{\Lambda}$,
a PPV of a finite system.
Here, 
we generalize the notion of the cluster property
to finite systems as follows.
Since $\Xi_{\Lambda}$ is not generally an energy eigenstate,
it may evolve with time in finite systems, and
we denote 
$\Xi_{\Lambda}$ after an time interval $t$ as $\Xi_{\Lambda}(t)$.
It becomes time-invariant in the limit of $\Lambda \to {\bf Z}^d$
because it approaches $\Xi_{{\bf Z}^d}$,
which is a time-invariant state.
Hence, 
if we introduce 
a time scale $T$, which will be 
taken sufficiently long, 
then 
$\nu_{T,\Lambda}:=
\inf_{0 \leq t \leq T} |(\Xi_{\Lambda}(t),m(x) \Xi_{\Lambda}(t))|
\to |\nu_{{\bf Z}^d}|$
for any $T > 0$.
For a positive number $\varepsilon$
 ($\leq 1$), 
we define
 an {\it $\varepsilon$-correlation region}  
$\Omega_{T, \Lambda}(y,\varepsilon)$ of a quantum state,
the expectation value for which is denoted by
$\langle \cdots \rangle$,
by its complement, 
\begin{eqnarray}
\Omega_{T,\Lambda}(y,\varepsilon)^{c}:=
\big\{&&x\in \Lambda \ \big|\ 
| \langle \delta a^*(t) \delta b(t) \rangle|
< \varepsilon 
(\langle \delta a^*(t) \delta a(t) \rangle)^{1/2}
(\langle \delta b^*(t) \delta b(t) \rangle)^{1/2},
\forall a 
\in {\cal A}(x), \forall b \in {\cal A}(y),
0 \leq \forall t \leq T
\big\},
\label{Omegac}\end{eqnarray}
where ${\cal A}(x)$ and ${\cal A}(y)$ denote a set of all
one-site operators (independent of $\Lambda$) 
at sites $x$ and $y$, respectively,
and $\delta a := a - \langle a \rangle$,
$\delta b := b - \langle b \rangle$.
We say that
a sequence of states (of finite systems) 
has a {\it cluster property} 
iff the correlation region $\Omega_{T,\Lambda}(y,\varepsilon)$ 
for any positive $\varepsilon$ and $T$
does not depend on $|\Lambda|$ 
for a sufficiently large $|\Lambda|$. 
Note that the volume $|\Omega_{T, \Lambda}(\varepsilon,y)|$ 
is independent of $y$ because of the assumed translational 
invariance of the states.
We thus denote it simply $|\Omega_{T, \Lambda}(\varepsilon)|$.
From Eq.\ (\ref{Omegac}), one can show 
for any one-site operator $a(x)$, 
and for $0 \leq \forall t \leq T$,
that 
\begin{equation}
(\Xi_{\Lambda}(t),
\delta A_{\Lambda}^* \delta A_{\Lambda} \Xi_{\Lambda}(t))
\leq \left(
|\Omega_{T,\Lambda}(\varepsilon)|/|\Lambda|
+\varepsilon
\right)
(\Xi_{\Lambda}(t),\delta a^*(0) \delta a(0) \Xi_{\Lambda}(t)),
\label{eqn:nofluct}
\end{equation}
where
$A_{\Lambda}$ 
is an intensive operator composed of $a$;
\begin{equation}
A_{\Lambda}:=\frac{1}{|\Lambda|}\sum_{x\in \Lambda}a(x),
\label{intensiveOp}\end{equation}
and 
$\delta A_{\Lambda} := A_{\Lambda} - \langle A_{\Lambda} \rangle$. 
By taking $\varepsilon$ small enough, one can see that 
fluctuations of any intensive operators are negligible
for PPVs, in consistent with thermodynamics.
In this sense, PPVs are thermodynamically normal.
For the order parameter
$M_{\Lambda}:=(1/|\Lambda|)\sum_{x \in \Lambda}m(x)$, 
in particular, 
$ 
(\Xi_{\Lambda}, \delta M_{\Lambda}^* \delta M_{\Lambda} \Xi_{\Lambda})
\to 0$
as $|\Lambda| \to \infty$.

For a finite system,
in general, there also exists the ground state
$\Phi_{0,\Lambda}$ which {\em preserves} the symmetry,
i.e., $(\Phi_{0,\Lambda}, M_{\Lambda} \Phi_{0,\Lambda})=0$
\cite{HL,miyashita,KT,SMpre}.
Since this state consists primarily of a superposition of
$\Xi_{\Lambda}$'s with different values of
$\nu_{\Lambda}$,
it has a large fluctuation of the order parameter
 \cite{HL,miyashita,KT,SMpre};
\begin{equation}
(\Phi_{0,\Lambda}, \delta M_{\Lambda}^* \delta M_{\Lambda} \Phi_{0,\Lambda})
={\cal O}(|\Lambda|^0), 
\label{abnormal}\end{equation}
which must be contrasted with that of PPVs. 
Since such a large fluctuation is anomalous in view of 
thermodynamics, 
we call the sequence of $\{\Phi_{0,\Lambda}\}$,
as well as its element $\Phi_{0,\Lambda}$,
the AFV.
%
It was proved that such a state 
cannot be a pure state
in the infinite-volume limit ($| \Lambda | \to \infty$)
\cite{ruelle,haag}.
In contrast to $\Xi_{\Lambda}$, 
$\Phi_{0,\Lambda}$ does not evolves with time in a closed system
because it is an eigenstate of $H_{\Lambda}$.

As an example, we consider a spin system with 
the simplest Hamiltonian,
$
H_{\Lambda}=- J \sum_{<x,y>} s_3(x) s_3(y),
$
which possesses a discrete symmetry,
the up-down symmetry.
The order parameter is 
$
S_{3,\Lambda} := (1/|\Lambda|)\sum_{x \in \Lambda}s_3(x).
$
There are two PPVs, 
$
\Xi_{+,\Lambda} := |+++\cdots\rangle
$
and
$
\Xi_{-,\Lambda} :=|---\cdots\rangle
$,
for which 
%
$ 
(\Xi_{\pm,\Lambda}, \delta a(0) \delta b(x)\Xi_{\pm,\Lambda})
=0
$
for $x \neq 0$.
Hence, $|\Omega_{0,\Lambda}(0)|=1$ for these states.
Since $\Xi_\pm$ are eigenstates of $H_{\Lambda}$ in this simple case, 
$|\Omega_{T,\Lambda}(0)|=|\Omega_{0,\Lambda}(0)|=1$ for any $T$.
On the other hand, 
$\Phi_{0,\Lambda} :=(\Xi_+ +\Xi_-)/{\sqrt{2}}
$
is an AFV \cite{degenerate}, for which 
$
(\Phi_{0,\Lambda},S_{3,\Lambda} \Phi_{0,\Lambda}) 
=
0
$
and
$
(\Phi_{0,\Lambda},S^*_{3,\Lambda} S_{3,\Lambda} \Phi_{0,\Lambda})
=
1
$.
As another example, we consider a free boson system
confined in a uniform box under the periodic boundary conditions.
The order parameter is
$m(x):=\psi(x)$.
The number state of free bosons $|N \rangle $, 
which is the number state of the lowest ($k=0$) single-body 
state, is the AFV because
$
\langle N|\delta M_{\Lambda}^* \delta M_{\Lambda} |N\rangle = N/|\Lambda|
=
{\cal O}(|\Lambda|^0)
$
when $N$ is increased in proportion to $|\Lambda|$.
On the other hand, 
the coherent state of free bosons $|\alpha \rangle$, 
which is the coherent state of the $k=0$ state, 
is easily shown to be a PPV.
Unlike these trivial examples, 
it is generally difficult to find PPVs and AFVs of interacting
many-body systems,  
and to confirm the cluster property of PPVs for {\em any} observables.
A successful example is 
interacting many bosons 
confined in a uniform box under the periodic boundary conditions
\cite{SMpre,SMcondmat,SMprl}.
It was shown that the  
`coherent state of interacting bosons' (CSIB) $| \alpha, G \rangle$
is a PPV \cite{SMcondmat}, 
which preserves the cluster property 
over $T = O(|\Lambda|^{1/2})$ \cite{SMpre}.
On the other hand, 
by superposing $| \alpha, G \rangle$'s over the phase of 
$\alpha$, one can construct the
`number state of interacting bosons' (NSIB) $| N, G \rangle$, 
which is the AFV 
\cite{SMcondmat}.
The fragility of the NSIB and the robustness of the CSIB 
were shown in Ref.\ \cite{SMprl}, in consistent with
the general theorems presented below.

We study robustness, against 
weak perturbations from a general environment E, 
of the AFV and PPVs of a general system
(which we hereafter call a `principal system') 
of size $|\Lambda|$.
The Hilbert space of the total system
is the product ${\cal H}_{\Lambda} \otimes {\cal H}_{\rm E}$
of the individual Hilbert spaces.
The Hamiltonian of the total system
is composed of three parts, 
$ 
H_{\rm tot}:=H_{\Lambda}+ H_{\rm int} + H_{\rm E},
$ 
where $H_{\Lambda}$ and $H_{\rm E}$ denote
the Hamiltonians of the principal system and E,
respectively, and 
$H_{\rm int}$ is an interaction between them.
The dimensionless constant $\lambda$ is small: 
when the principal system couples strongly to a part of 
an external system, one must include such a part into the principal system,
then, after a proper renormalization process, 
the principal system  
couples only weakly to the rest of the external system, which we call here 
the environment.
Most previous work on decoherence of macroscopic systems
assumed that 
the principal system could be described 
by a {\it small number} of collective coordinates,
which interact {\em non-locally} 
with some specific environment.
As mentioned in the introduction, however, such 
formulations are inappropriate for general systems.
Therefore, we start from a Hamiltonian $H_{\rm tot}$ with
{\em macroscopically large} degrees of freedom, which 
interact {\em locally}
with many degrees of freedom of E; 
\begin{eqnarray}
H_{\rm int}=\lambda \sum_{x \in \Lambda_{\rm C}} a(x)\otimes  b(x),
\label{eqn:int}
\end{eqnarray}
where $a(x)$ and $b(x)$ are local operators of the principal system
and E, respectively,  
at the lattice point $x \in \Lambda_{\rm C}$.
Here, $\Lambda_{\rm C}$ ($\subseteq \Lambda$) is a `contact region' 
between the principal system and E.
Without loss of generality, 
we assume that 
$
\mbox{tr}(\sigma b(x)) =0
$:
If it is finite 
it can be absorbed into $H_{\rm tot}$ 
as a renormalization term.
Putting
$
a_k:
=
|\Lambda|^{-1} \sum_{x\in \Lambda}
a(x) \mbox{e}^{ikx} 
$and
$
b_k:
=
\sum_{x\in \Lambda_{\rm C}}
b(x) \mbox{e}^{-ikx}
$,
for $k \in (2\pi {\bf Z} / L)^d$,
 yields
$
H_{\rm int} =\sum_{k} a_k \otimes b_k
$ \cite{b0g00}.

The density operator of the total system 
$\rho_{\rm tot}(t)$ evolves according to $H_{\rm tot}$.
We are interested in a reduced density operator of
the principal system,
$
\rho(t)
:=\mbox{tr}_{{\cal H}_{\rm E}}[\rho_{\rm tot}(t)].
$
Since we discuss decoherence of an initially pure state, 
we assume that $\rho_{\rm tot}$ is initially an uncorrelated product;
$
\rho_{\rm tot}(0)
=
\rho(0) \otimes \sigma
$,
where 
$\rho(0) := |\phi \rangle \langle \phi|$ is a pure state of 
the principal system, 
and $\sigma$ is a time invariant state
of E.
We are studying two cases: 
$\rho(0)$ is (a) the AFV, and (b) a PPV.
In the former case $\rho(0)$ is an eigenstate
of $H_{\Lambda}$,
whereas in the latter $\rho(0)$ is a superposition of 
low-lying eigenstates of $H_{\Lambda}$  \cite{HL,miyashita,KT,SMpre}.
In either case, the energy spread 
(the width of distribution over eigenvalues of 
$H_{\Lambda}$) of 
$\rho(0)$ is narrow.
It is expected that 
for a sufficiently small $\lambda$
the energy spread remains small 
for a short $t$ of interest.
If the energy spread is smaller than $\hbar/\tau_c$, where
 $\tau_c$ denotes the correlation time of $b$ of E, 
we obtain the following Markovian equation;
\begin{eqnarray}
i \hbar \frac{d}{dt}\rho =
[H_{\Lambda},\rho] 
+
i \frac{\lambda^2}{\hbar}\sum_{k_1}\sum_{k_2} g_{k_1 k_2}
(
2 a_{k_2} \rho a^*_{k_1}
-\{a^{*}_{k_1}a_{k_2}, \rho \} 
),
\label{eqn:master}
\end{eqnarray}
where $g$ is a positive matrix defined by
the time correlation in E;
\begin{eqnarray}
g_{k_1 k_2} 
:
=\frac{1}{2}
\int^{\infty}_{-\infty}ds \langle b_{k_1}^* b_{k_2}(s) \rangle.
\end{eqnarray}

Due to perturbations from E, 
initially pure states generally evolve into mixed ones.
If a pure state rapidly evolves 
into a mixed state (i.e., decoheres), such a state
should be hard to realize and observe.
We say such states  are `fragile.'
On the other hand, 
if a pure state does not decohere for a long time, 
it should be easy to observe.
We say such states  are `robust.'
Namely, effects of the environment 
select out particular states as observable ones.
Zurek et al.\ called this mechanism the 
`environment-induced superselection rule'
in his discussion on 
quantum measurements \cite{Zurek}.
We apply this idea to the present problem of SSB in a finite system.
As a measure of purity of a quantum state, 
we employ the so-called 
linear entropy
$
S_{\rm lin}(\rho):=1- \mbox{tr}[\rho^2]
$  \cite{entropy},
which vanishes only for pure states.
We evaluate $S_{\rm lin}$ as a power series of $\lambda^2$,
$S_{\rm lin} = S_{\rm lin}^{(0)} + S_{\rm lin}^{(1)} + \cdots$, 
where $S_{\rm lin}^{(n)} = {\cal O}(\lambda^{2n})$,
using the standard interaction picture technique.
We confirmed that this series 
converges \cite{MS}.
Since $S_{\rm lin}^{(0)}=0$ for $\rho(0) = |\phi \rangle \langle \phi|$, 
this suggests that 
$S_{\rm lin}^{(1)}$ would give the dominant contribution
under our assumption that $\lambda$ is small.
It is calculated as
\begin{eqnarray}
S_{\rm lin}^{(1)}(\phi,t)
=
\frac{\lambda^2}{\hbar^2}\int^{t}_0
ds 
\sum_{k_1 k_2} g_{k_1 k_2}
(\phi, \delta a_{k_1}^*(s) \delta a_{k_2}(s)\phi).
\label{eqn:1stlin}
\end{eqnarray}
If $\phi$ is translational invariant, 
both spatially and temporally, 
the rhs is bounded by the fluctuation
of an intensive variable $A_{\Lambda}:=(1/|\Lambda|)\sum_{x\in \Lambda}a(x)$, 
and we obtain
\\{\it
Theorem 1:}
\begin{equation}
S_{\rm lin}^{(1)}(\phi,t)
\geq
(\lambda^2/\hbar^2) g_{00}
(\phi, \delta A_{\Lambda}^* \delta A_{\Lambda} \phi) t.
\label{th1}\end{equation}
Since the rhs is proportional to $t$,
we may interpret it divided by $t$ as
a {\it lower bound of the decoherence rate}, 
which we denote $\gamma$.
It is proportional to 
the fluctuation of the intensive variable $A$ composed of 
$a(x)$ which constitutes $H_{\rm int}$ as Eq.\ (\ref{eqn:int}).

To apply this theorem,
recall that we are considering a theory which 
effectively describes  
phenomena in some energy range of interest.
The effective theory can be constructed from 
an elementary dynamics by an appropriate renormalization 
process.
In this process, in general, 
many terms would be generated in 
the effective interaction;
$
H_{\rm int}
=
H_{\rm int}^{[1]} + H_{\rm int}^{[2]} + \cdots,
$
where 
$
H_{\rm int}^{[\ell]}
=
\lambda^{[\ell]} \sum_{x \in \Lambda_{\rm C}^{[\ell]}} a^{[\ell]}(x)
\otimes  b^{[\ell]}(x)$
 \cite{Hlocal}.
Hence, it seems rare that 
$H_{\rm int}$ does not have a term with 
$a^{[\ell]}(x) = m(x)$,
although $\lambda^{[\ell]}$ might be small.
If $\lambda^{[\ell]}$ is small, 
such a term could be neglected if $|\Lambda|$ were small.
However, it becomes relevant in the present case of $|\Lambda| \gg 1$, 
for the following reason.
Such a term yields
$
\gamma^{[\ell]} = (\lambda^{[\ell] 2}/\hbar^2) g_{00}^{[\ell]}
\times {\cal O}(|\Lambda|^0)
$
for $\gamma$ of the AFV.
For PPVs, on the other hand, 
we can see from Eqs.\ (\ref{eqn:nofluct}) and (\ref{th1}) that
$ 
\gamma^{[\ell]} = (\lambda^{[\ell] 2}/\hbar^2) g_{00}^{[\ell]}
\times {\cal O}(1/|\Lambda|)
$
for {\em any} of $\hat H_{\rm int}^{[\ell]}$'s.
Since $|\Lambda|$ is a macroscopic number, 
the former is much larger than the latter.
Regarding the factor $g_{00}$, 
we can estimate its order of magnitude as follows \cite{b0g00}.
Let 
$\Lambda_{\rm E}^{\rm corr}$ be the {\it correlation region of E}, i.e., 
the region of $x$ in which 
$\int^{\infty}_{-\infty}ds \langle b^{*}(x) b(0,s) \rangle$
is correlated.
When $|\Lambda_{\rm E}^{\rm corr}| > |\Lambda_{\rm C}|$, 
we can roughly estimate that $g_{00} \propto |\Lambda_{\rm C}|^2$ \cite{MS}.
Hence, Theorem 1 yields
$\gamma \propto (\lambda^2/\hbar^2) |\Lambda_{\rm C}|^2$
for the AFV \cite{ekert}, 
whereas
$\gamma \propto (\lambda^2/\hbar^2) 
|\Lambda_{\rm C}|^2/|\Lambda|$ for PPVs \cite{Th1PPV}.
On the other hand, 
when $|\Lambda_{\rm E}^{\rm corr}| < |\Lambda_{\rm C}|$, 
we can roughly estimate that 
$g_{00} \propto |\Lambda_{\rm C}||\Lambda_{\rm E}^{\rm corr}|$.
Hence, 
$\gamma \propto (\lambda^2/\hbar^2) 
|\Lambda_{\rm C}||\Lambda_{\rm E}^{\rm corr}|$ for the AFV, 
whereas
$\gamma \propto (\lambda^2/\hbar^2) 
|\Lambda_{\rm C}||\Lambda_{\rm E}^{\rm corr}|/|\Lambda|$ for PPVs.
In both cases, we find that 
the AFV is fragile (i.e., decoheres at an anomalously 
fast rate),
however small $\lambda$ is, 
if $|\Lambda_{\rm C}|$ ($\leq |\Lambda|$) is large enough.
Therefore, 
we think that 
AFVs are almost always fragile in real physical 
systems.
This seems to give microscopic foundations of our experience;
AFVs are difficult to observe.

How are PPVs?
We have already seen that $\gamma$
is ${\cal O}(1/|\Lambda|)$ times 
smaller for PPVs than for the AFV.
Unlike the case of Ref.\ \cite{SMprl}, however, 
we cannot draw a general conclusion on the robustness of PPVs
because $\gamma$ is a lower bound.
To see more details, 
we now present another theorem.
We can prove it for two cases \cite{MS}; 
the breaking of (a) the ${\bf Z}_2$ (parity) symmetry, 
and (b) the $U(1)$ symmetry, 
under the assumption that 
$a(x) = m(x)$ in Eq.\ (\ref{eqn:int}).
We here describe an outline of the proof for case (a).
In this case, 
$m(x)$ transforms as ${\cal P} m(x) {\cal P}^\dagger = - m(x)$ 
by the parity operation ${\cal P}$.
Any vector can be decomposed into 
even- and odd-parity components, 
$\Phi_+$ and $\Phi_-$, respectively.
Since the Hamiltonian commutes with ${\cal P}$, 
they remain in the 
even- and odd-parity subspaces, respectively, for any $t$.
We therefore denote them $\Phi_+(t)$ and $\Phi_-(t)$, 
which are assumed to be normalized.
Let us consider a PPV which can be decomposed as
$ 
\Xi_{\Lambda}(t):=c_+ \Phi_+(t) +c_- \Phi_-(t).
$ 
By operating ${\cal P}$, we obtain another PPV, 
$\Xi'_{\Lambda}(t):=  {\cal P} \Xi_{\Lambda}(t) =
c_+ \Phi_+(t) -c_- \Phi_-(t)$.
Since 
$\Xi_{\Lambda}$ and $\Xi'_{\Lambda}$ must become orthogonal to each other
when $|\Lambda| \to \infty$, 
we obtain
$
c_+,c_- \to 1/\sqrt{2}
$
as $|\Lambda| \to \infty$.
For a macroscopic time region $0\leq t \leq T$, 
we denote the $\varepsilon$-correlation region of $\Xi_{\Lambda}$
by $\Omega_{T,\Lambda}(\varepsilon)$, 
and let
$
\nu_{T,\Lambda} :=\mbox{inf}_{0\leq t \leq T}
|(\Xi_{\Lambda}(t),M_{\Lambda} \Xi_{\Lambda}(t))|
$.
After lengthy calculations, we obtain
\begin{eqnarray}
|c_+|^2 S_{\rm lin}^{(1)}(\Phi_+,t)+|c_-|^2 S_{\rm lin}^{(1)}(\Phi_-,t)
-S_{\rm lin}^{(1)}(\Xi_{\Lambda},t)
\geq
\frac{\lambda^2}{\hbar^2}
g_{00}
\Big\{
\nu_{T,\Lambda}^2 t - 
\left[
\frac{|\Omega_{T,\Lambda}(\varepsilon)|}{|\Lambda|}+\varepsilon
\right]
\int^t_0 ds 
(\Xi_{\Lambda}(s), \delta m^*(0) \delta m(0) \Xi_{\Lambda}(s))
\Big\}.
\nonumber
\end{eqnarray}
Noting that any local operator
hardly intertwine between $\Xi_{\Lambda}$ and 
 $\Xi'_{\Lambda}$, we can show that
 $
 |c_+|^2 S_{\rm lin}^{(1)}(\Phi_+,t)+|c_-|^2 S_{\rm lin}^{(1)}(\Phi_-,t)
 \to
 S_{\rm lin}^{(1)}(\Phi_{0,\Lambda})
 $
 as $\Lambda \to {\bf Z}^d$, 
 and that
 $\nu_{T,\Lambda}^2 
 \geq
 (\Phi_{0,\Lambda}, \delta M_{\Lambda}^* \delta M_{\Lambda} \Phi_{0,\Lambda})
 +\epsilon'_{\Lambda}
 $
 with $\epsilon'_{\Lambda} \to 0$.
We thus obtain
\\ {\it
Theorem 2 : 
For a fixed contact region $\Lambda_{\rm C}$},
\begin{eqnarray}
 S_{\rm lin}^{(1)}(\Phi_{0,\Lambda},t)
-S_{\rm lin}^{(1)}(\Xi_{\Lambda},t)
\geq
\frac{\lambda^2}{\hbar^2}g_{00} t 
(\Phi_{0,\Lambda},\delta M_{\Lambda}^* \delta M_{\Lambda} \Phi_{0,\Lambda})
+\epsilon_{\Lambda}.
\label{th2}
\end{eqnarray}
{\it where $\epsilon_{\Lambda}$ is a small number, approaching $0$ 
as $\Lambda \to {\bf Z}^d$}.\\ 
Since the rhs is proportional to $t$,
we can interpret it divided by $t$ as 
a lower bound of the {\it difference} of the decoherence rates, 
which we denote $\Delta \gamma$. 
In a manner similar to the estimation of
$\gamma$, 
we can roughly estimate that 
$\Delta \gamma \propto (\lambda^2/\hbar^2) |\Lambda_{\rm C}|^2$
for $|\Lambda_{\rm E}^{\rm corr}| > |\Lambda_{\rm C}|$,
whereas
$\Delta \gamma \propto (\lambda^2/\hbar^2) 
|\Lambda_{\rm C}||\Lambda_{\rm E}^{\rm corr}|$
for $|\Lambda_{\rm E}^{\rm corr}| < |\Lambda_{\rm C}|$.
In both cases, we find that 
$\Delta \gamma$ is large,
however small $\lambda$ is, 
if $|\Lambda_{\rm C}|$ ($\leq |\Lambda|$) is large enough.
This large term originates from 
the modes with $k=0$ \cite{b0g00}, 
whereas Theorem 2 indicates that the other modes with $k\neq0$
give only a negligible difference.
Namely, both the AFV and PPVs decohere by the $k \neq 0$ modes,
whereas only the AFV decohere 
anomalously fast by the $k=0$ modes.
In some cases
(e.g., when the environment is violent)
the former modes might make both the AFV and PPVs decohere quickly.
Hence, unlike the case of Ref.\ \cite{SMprl}, 
we cannot draw a definite conclusion on the robustness of PPVs
for general cases.
We can, however, definitely say that 
PPVs are less fragile than the AFV in the sense that 
$
 S_{\rm lin}^{(1)}(\Phi_{0,\Lambda},t)
-S_{\rm lin}^{(1)}(\Xi_{\Lambda},t)
\geq 0
$,
because the rhs of Eq.\ (\ref{th2}) is positive.

To demonstrate how the theorems are satisfied, we present 
simple examples, for which we can explicitly calculate $S_{\rm lin}^{(1)}$.
For the simple spin system discussed above, 
one can easily show by putting $m(x):=s_3(x)$ that 
$a_k \Xi_+ =\delta_{k 0} \Xi_+$ and so on, and that 
$S_{\rm lin}^{(1)}(\Phi_{0,\Lambda},t)=(\lambda^2/\hbar^2)g_{00} t$, 
$S_{\rm lin}^{(1)}(\Xi_+,t)= 0$.
For the free boson system, on the other hand, 
if we assume $H_{\rm int}:=\sum_{x}( \psi(x) \otimes b(x)
+\psi^*(x) \otimes b^*(x) )$, 
we can show that 
$
S_{\rm lin}^{(1)}(|N \rangle, t)
= (\lambda^2/\hbar^2)
[ n_0 (g^{+}_{00} +g^-_{00})
+\sum_{k}
g^{-}_{kk}/|\Lambda|]t
$,
$
S_{\rm lin}^{(1)}(|\alpha \rangle, t)
= (\lambda^2/\hbar^2)
[\sum_{k}
g^{-}_{kk}/|\Lambda|]t, 
$
where $g^+_{00},g^-_{kk}$ are constants determined by 
correlation functions in E.


TM thanks H.\ Kato for helpful discussions.



\begin{references}
\bibitem{entropy}
Although the linear entropy does not have some
good properties of other entropies, it 
equals to the Renyi entropy, 
$
S_2[\rho]=-\mbox{ln}(\mbox{tr}\rho^2)
$, to ${\cal O}(\lambda^2)$.
\bibitem{degenerate}
Since this example is too simple, the ground states (AFVs) are degenerate.
\bibitem{Hlocal}
We expect that by an appropriate renormalization process 
$H_{\rm tot}$ can be made local in the relevant space-time
scale.
\bibitem{Th1PPV}
Although Theorem 1 was derived for time invariant states,
it can also be applied to PPVs for 
a sufficiently large $|\Lambda|$, because for which PPVs are time invariant.
\end{references}
\end{document}